\documentclass[prb,twocolumn,showpacs]{revtex4}
\usepackage{amsfonts}
\usepackage{amsmath}
\usepackage{amssymb}
\usepackage{graphicx}

\begin{document}

\title{Precision determination of band offsets in strained InGaAs/GaAs
quantum wells by C-V-profiling and Schr\"{o}dinger-Poisson
self-consistent simulation}
\author{V.~I.~Zubkov}
\author{M.~A.~Melnik}
\author{A.~V.~Solomonov}
\author{E.~O.~Tsvelev}
\affiliation{St.\ Petersburg State Electrotechnical University,
Prof.\ Popov str.\ 5, 197376, St.-Petersburg, Russia}
\author{F.~Bugge}
\author{M.~Weyers}
\author{G.~Tr\"{a}nkle}
\affiliation{Ferdinand-Braun-Institut f\"{u}r
H\"{o}chstfrequenztechnik, Albert-Einstein-Str.\ 11, D-12489
Berlin, Germany}

\keywords{} \pacs{73.21.Fg, 81.07.St}

\begin{abstract}
The results of measurements and numerical simulation of charge
carrier distribution and energy states in strained quantum wells
In$_x$Ga$_{1-x}$As/GaAs ($0.06 \leqslant x \leqslant 0.29$) by
C-V-profiling are presented. Precise values of conduction band
offsets for these pseudomorphic QWs have been obtained by means of
self-consistent solution of Schr\"{o}dinger and Poisson equations
and following fitting to experimental data. For the conduction
band offsets in strained In$_x$Ga$_{1-x}$As/GaAs - QWs the
expression ${\Delta}E_{C}(x) = 0.814x - 0.21x^{2}$ has been
obtained.
\end{abstract}

\date{\today}
\maketitle

\section{Introduction}
Since the development of semiconductor heterostructures the
determination of energy band discontinuities of various
semiconductor pairs has been a very important task. Energy band
offsets dominantly control the electronic states in
heterostructures and, hence, the output parameters of
semiconductor devices. The importance of getting true values of
band offsets as well as the difficulties in obtaining and, even
more, in interpreting the relevant data have been attracting
attention for the last 30 years. R. Dingle was one of the first
who reported in 1974-75 \cite{Din74,Din75} the value of band
offsets for isoperiodic heterosystem (Al-Ga)As/GaAs ("Dingle" rule
85:15). Then H. Kroemer, \cite{Kro83,Kro84,Kro86} G. Duggan
\cite{Dug85} and Yu et al.\ \cite{Yu92} comprehensively reviewed
the understanding of band offsets before 1991 and provided an
overview of the methods commonly used in experimental band offset
determination, mostly optical at that time. At the same time, the
authors \cite{Jog90} and others showed that a low sensitivity of
the optical transition energies to the band offsets made its
determinations rather confusing. Up to now a great number of
papers has been published on this subject (see bibliography in
recent comprehensive review \cite{Vur01}). So far, however, as was
pointed out in the review, among the ternary alloys used in
quantum electronics, only the AlGaAs/GaAs system has generally
accepted values of band offsets.

For one of the most important used heteropairs --
In$_x$Ga$_{1-x}$As/GaAs -- as yet no clear picture about the
dependence of band offsets on alloy composition has been obtained,
despite the very intensive investigations in last years. The data
collected by P. Bhattacharya \cite{Bha93} show a great scatter of
the values of relative conduction band offset ${\Delta}E_{C}$
between 35\% and 85\% for $x <$ 0.35. Above mentioned review
\cite{Vur01} reports relative conduction band offsets for the
In$_x$Ga$_{1-x}$As/GaAs -- system in the range 57--90\% and
recommends as a rule of thumb ${\Delta}E_C[\text{eV}] = x$ for $x
<$ 0.5. They conclude that no detailed study has yet been carried
out on InGaAs-based heterojunctions. Recent publications on this
subject \cite{Ari00,Dis97,Kar96,Liw98} only present partial
results for different compositions, more or less agreed with
"recommended" in Ref.~\onlinecite{Vur01}. Theoretic calculations
\cite{Wei98} give the valence band offset for the end combination
InAs/GaAs ${\Delta}E_V = 0.06$ eV, which is in serious
disagreement with experimental data.

One important device application of the heterosystem InGaAs/GaAs
is high power laser diodes with strained quantum wells.
\cite{Bug98} In these structures thin quantum-size layers of
InGaAs grow pseudomorphically, i.e. having the lattice constant of
the underlying GaAs-layer in the plane of the heterojunction. The
elastic energy, accumulated due to crystal cell distortion, causes
the band structure of the thin InGaAs layer to be modified,
\cite{Dav98} altering particularly its energy gap. Hence, in
strained InGaAs/GaAs quantum wells one should expect another band
offsets than in heterostructures with thick layers of the solid
solution. In cases between pseudomorphic growth and full strain
relaxation (occurs in thick layers) the band offset in InGaAs/GaAs
will have, obviously, some intermediate values. This fact
explains, we suppose, the variety of data found in literature.

Numerical fitting of C-V-curves \cite{Let91, Sub93, Tsc96} by
means of self-consistent solution of Schr\"{o}dinger and Poisson
equations is one of the most promising approaches to measure band
offsets of quantum well structures. \cite{Tan90, Bro96, Ari00}
This approach correctly takes the quantization of carriers in a
quantum well into consideration and yields very accurate results.
However, well-defined heterostructures are necessary for this.

Complex multilayer structures like multiQWs etc.\ and unknown
dopant profile or the presence of deep levels add sources of
uncertainties. Therefore, in order to be sure to get precise
values for band offsets at heterojunctions simple structures with
a minimum of unknown parameters or parameters to be fitted should
be used.

This work presents accurate data for band offsets in
heterostructures with strained pseudomorphic
In$_x$Ga$_{1-x}$As/GaAs ($0 < x < 0.3$) quantum wells. To obtain
these values we have carried out a systematic cycle of
C-V-measurements on specially fabricated structures. Details of
sample preparation and measurements are described in Section
\ref{Sec:Samples}. In Section \ref{Sec:Model} the model for
simulating measured concentration profiles and deriving the values
of conduction band offsets based on self-consistent numerical
solution of the Schr\"{o}dinger and Poisson equations is
described. The carrier concentration in the quantum well region is
calculated on the base of a quantum-mechanical approach.
Mathematical aspects of the computations are presented in Sections
\ref{Sec:Poisson} and \ref{Sec:Schrodinger}. To increase the
accuracy of the numerical calculations a non-uniform mesh with the
mesh step inside the quantum well 10 times smaller than in the
other regions has been used. Finally, in Section \ref{Sec:Results}
we present the results of numerical fitting of experimentally
measured C-V curves. The dependence of conduction band offset for
strained pseudomorphically grown In$_{x}$Ga$_{1-x}$As/GaAs - QWs
has been obtained as a function of quantum well composition in the
range $0.06 \leqslant x \leqslant 0.29$.

\section{\label{Sec:Samples} Sample structure and measurement procedure}
\begin{figure}
\includegraphics [width=6.5 cm] {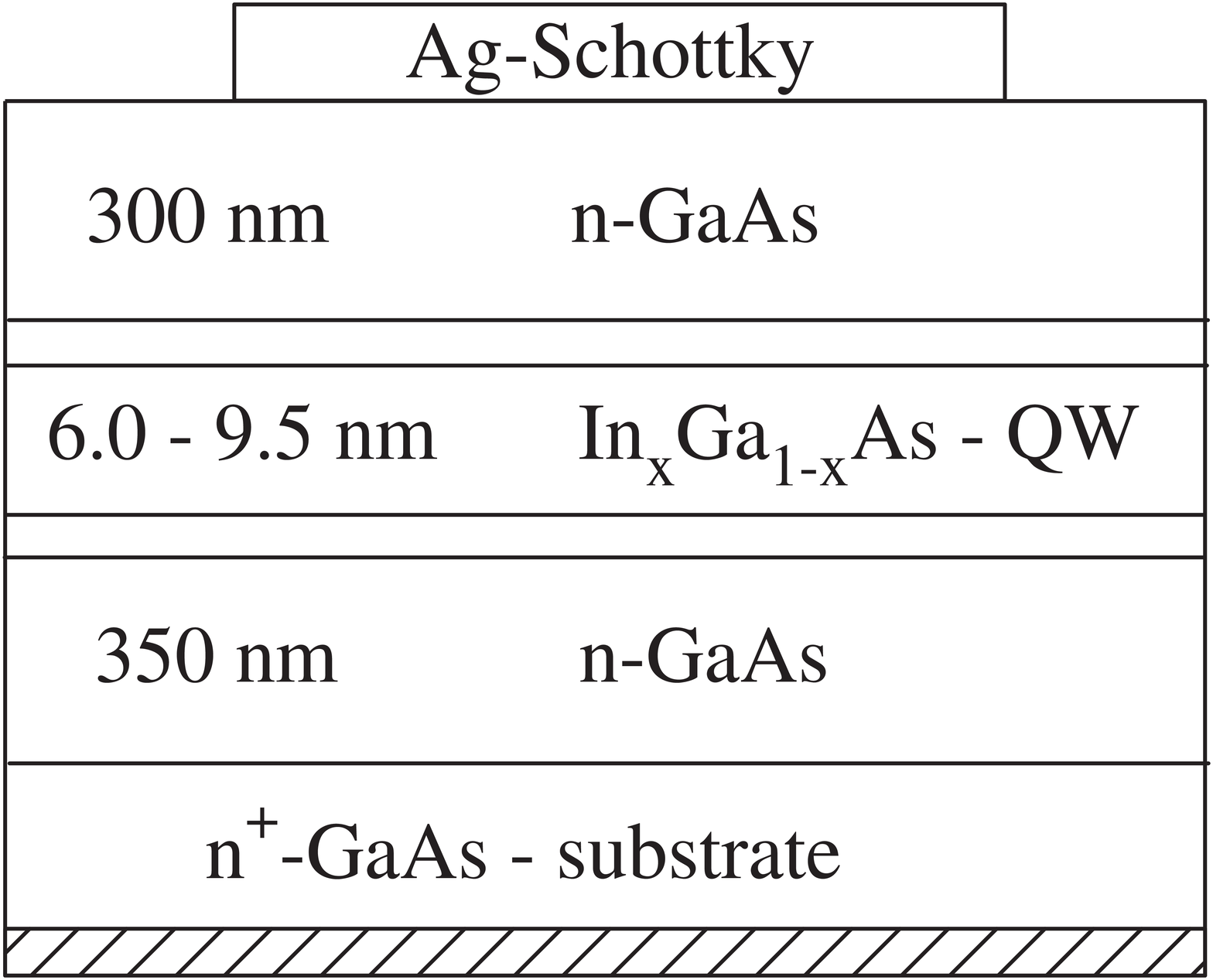}
\caption{\label{Fig:Sample} Layer sequence of the grown samples
with In$_{x}$Ga$_{1-x}$As/GaAs -- quantum wells.}
\end{figure}

A special set of high quality samples with a simplified structure
(Fig.~\ref{Fig:Sample}) containing In$_{x}$Ga$_{1-x}$As/GaAs
quantum wells of different width ($w=6.0-9.5$ nm) and composition
($x = 0.065-0.29$) was grown on $n^+$-GaAs substrates by MOVPE at
deposition temperatures of 650$^{\circ}$C and 770$^{\circ}$C. The
GaAs cladding layers were uniformly doped with Si, except for the
QWs themselves and thin (5 nm) spacer layers on both sides of the
quantum well. To get the best experimental results and to
eliminate possible uncertainties in the subsequent numerical
fitting, the cap GaAs layer was designed to be 300 nm thick and
have a constant doping level of $(6-7)*10^{16}$ cm$^{-3}$. The
width and composition of the QWs and cladding layers have been
determined by high resolution X-ray diffraction (HRXRD). All QWs
were fully strained without any relaxation seen in X-ray area
maps. \cite{Bug98} Ag-Schottky barriers were fabricated on top of
the structures and Ohmic contacts were formed on the substrate.

The parameters of the grown structures are listed in Table
\ref{tab:Samples}.
\begin{table*}
\caption{\label{tab:Samples} Characteristics of the structures and
results of numerical simulation of conduction band discontinuities
in In$_{x}$Ga$_{1-x}$As/GaAs strained quantum wells grown by
MOVPE.}
\begin{ruledtabular}
\begin{tabular}{ccccccc}
Sample & x & T deposition & d cap layer & QW width & E of bound level & $\Delta$E$_C$\\
\# &  & ($^{\circ}$C) & (${\mu}$m) & (nm) &  at 0V (meV) & (meV) \\
\hline
298 & 0.065 & 770 & 0.304 & 9.5 & -10.9 & 57\\
299 & 0.14 & 770 & 0.304 & 8.0 & -29.0  & 110\\
308 & 0.145 & 650 & 0.302 & 6.0 & -25.0  & 110\\
303 & 0.145 & 650 & 0.305 & 7.5 & -32.1  & 120\\
309 & 0.145 & 650 & 0.302 & 9.5 & -35.1  & 120\\
296 & 0.19 & 770 & 0.295 & 6.5 & -38.3  & 150\\
297 & 0.2 & 650 & 0.298 & 6.5 & -39.5 & 155\\
306 & 0.215 & 770 & 0.304 & 7.2 & -43.7 & 160\\
307 & 0.225 & 770 & 0.308 & 7.4 & -48.8 & 175\\
300 & 0.23 & 770 & 0.304 & 7.2 & -48.5  & 175\\
301 & 0.27 & 770 & 0.300 & 6.5 & -54.6  & 210\\
305 & 0.29 & 650 & 0.300 & 6.0 & -55.3 & 220\\
\end{tabular}
\end{ruledtabular}
\end{table*}

The measurements of capacitance-voltage characteristics and
profiling of majority carriers in the quantum wells have been
carried out with the help of a computer-controlled
C-V-profilometer at a testing frequency of 1 MHz and with an
amplitude of the probing signal of 15 or 50 mV.

\begin{figure}
\includegraphics[width=8.5cm]{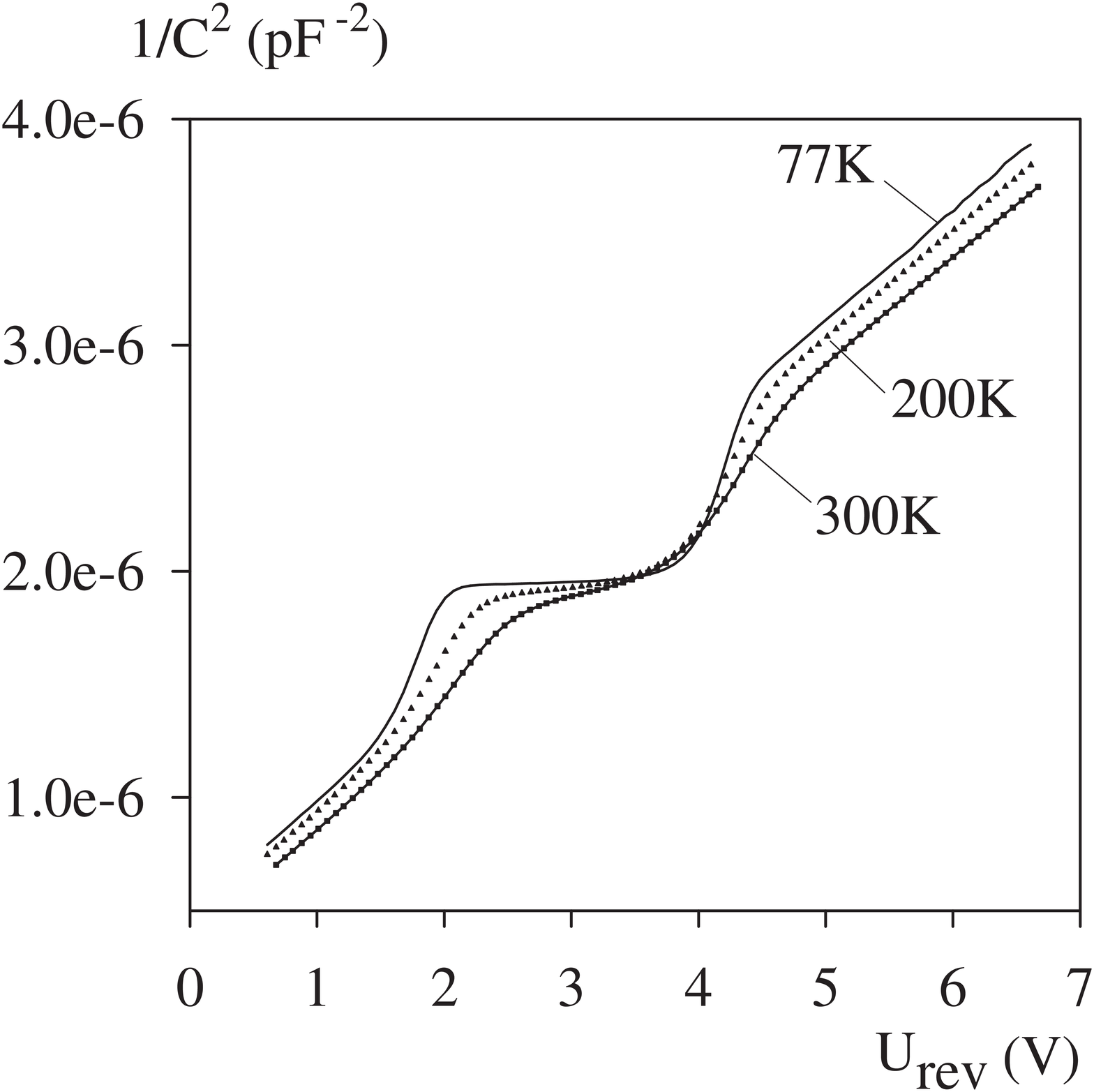}
\caption{\label{Fig:Cvinga} C-V-characteristics of
In$_{0.225}$Ga$_{0.775}$As/GaAs QW at different temperatures
(Sample \#307).}
\end{figure}

At zero bias the width of the space charge region under the
Schottky-barrier in the samples was less than the thickness of the
cap GaAs-layer. With increasing reverse bias the space charge
region was broadened and its border crossed the quantum well. The
C-V-characteristics of all samples clearly exhibit a plateau in
the range of $U_{\text{rev}} = 2-4.5$ V related to discharging
carriers in the QW. A typical example of $1/C^{2} =
f(U_{\text{rev}})$ characteristic for sample \#307 at different
temperatures is shown in Fig.~\ref{Fig:Cvinga}.

The apparent carrier distribution is derived from a measured
C-V-curve using the well known formula for depletion approximation
\begin{equation}\label{n_w}
n(d)=2\left[\epsilon\epsilon_0eA^2\frac{d}{dV}\left(\frac{1}{C^2}
\right)\right]^{-1},
\end{equation}
where $\epsilon$ is the dielectric constant (assumed to be equal
for both the well and the barriers), $e$ is the electron charge,
$A$ is the area of the Schottky diode. The depletion width $d$ is
given as usual by
\begin{equation}\label{w}
d=\frac{\epsilon\epsilon_0A}{C}.
\end{equation}

\begin{figure}
\includegraphics[width=8.5cm]{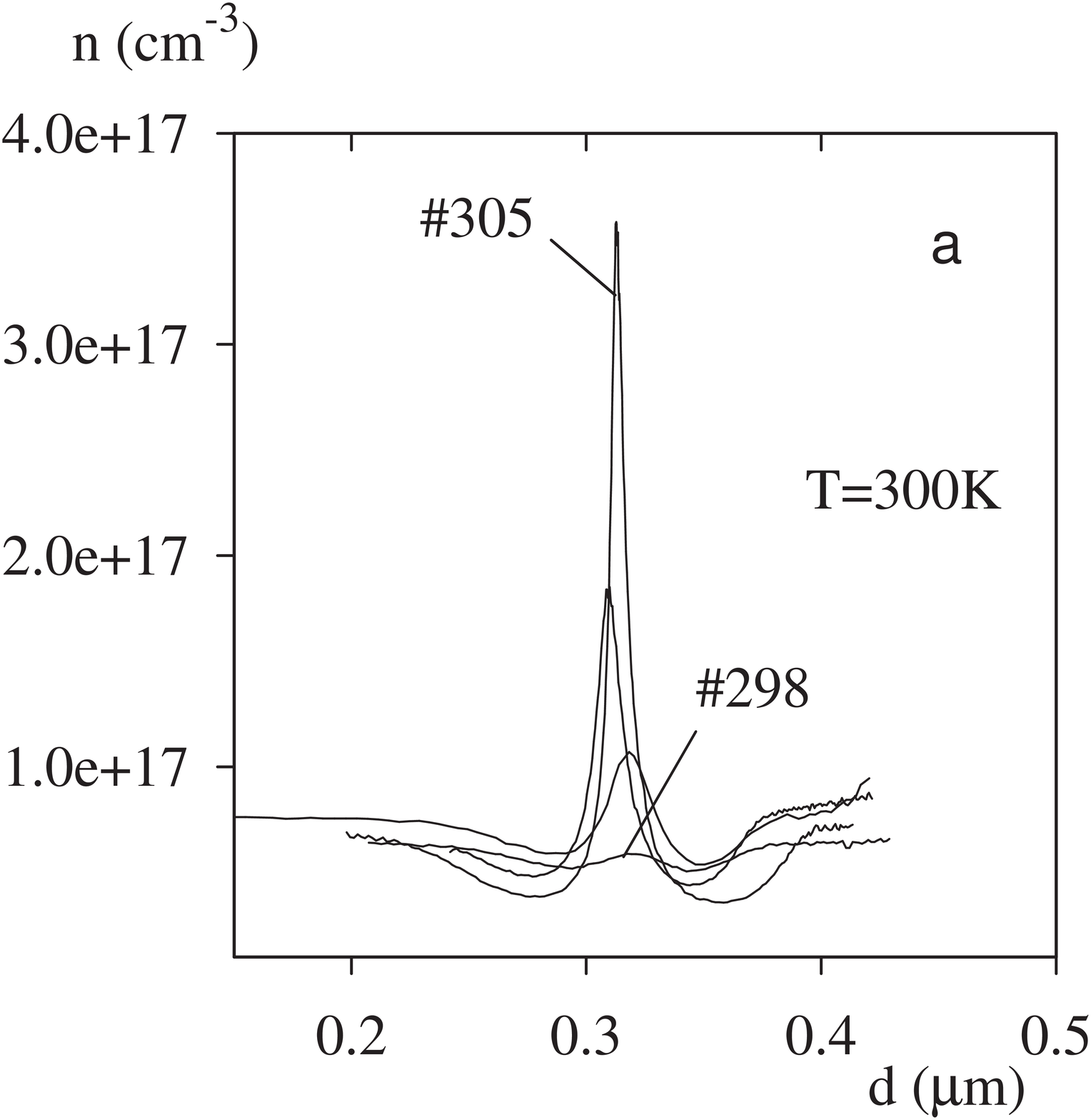}\\
\includegraphics[width=7.5cm]{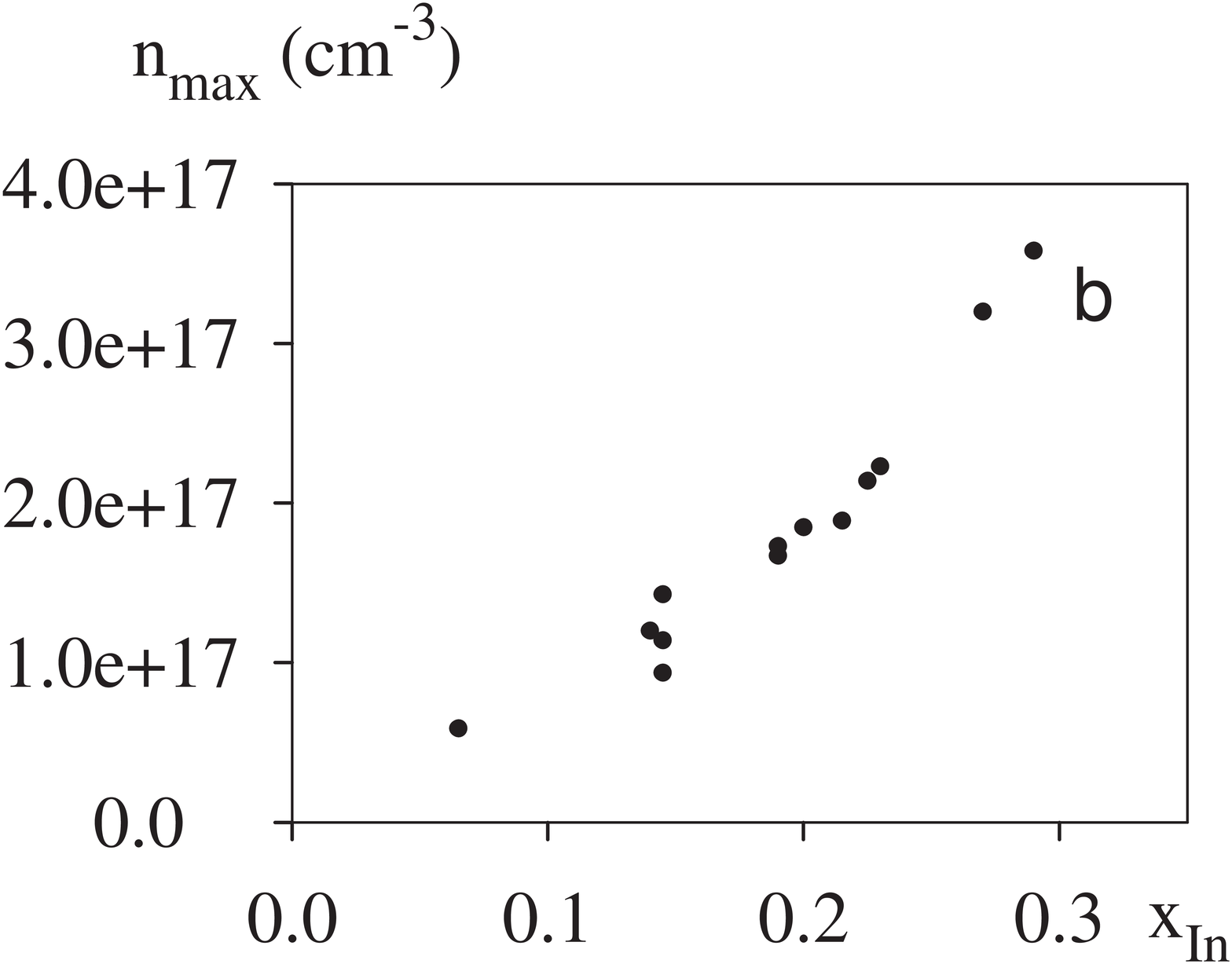}\\
\caption{\label{Fig:Common} C-V-profiling of
In$_{x}$Ga$_{1-x}$As/GaAs quantum wells. Samples \#296--\#309. a)
Common view of apparent carrier concentrations for several
samples, T=300K; b) Apparent concentration peak values as a function of $x$.\\
Note that near $x=0.14$ there are 4 samples with different width
of QW.}
\end{figure}

Figure ~\ref{Fig:Common} shows some examples of the $n-d$ curves
covering the whole range of QW compositions. The profiles
exhibited a clear dependence of amplitude, width, and the depth of
depletion on the composition and the width of the QWs. It is worth
to note that beyond the regions of accumulation and depletion
related to the QW the carrier concentration was excellently
constant, and we used it in the fitting of the experimental
profiles to the simulated ones on the base of self-consistent
solution of the Schr\"{o}dinger and Poisson equations by varying
the band offset.

\section{\label{Sec:Model} Model for simulating C-V profiles}

For simulating the C-V-characteristics the Poisson equation
\begin{equation}\label{Poisson}
\frac{d}{dz}\left(\epsilon_0\epsilon(z)\frac{d\varphi(z)}{dz}\right)=e\left[
N_D^+(z)-n(z)\right]
\end{equation}
has been solved, where $\varphi(z)$ is the electrostatic
potential, $n(z)$ is the free carrier concentration, and
$N_{D}^{+}$ is the concentration of ionized donors. The boundary
conditions for (\ref{Poisson}) at the Schottky barrier and in the
electroneutrality region, far away from the QW, are:
\begin{equation}\label{bound_Poiss}
\varphi(0)=U+\varphi_{bi},
\end{equation}
\begin{equation}\label{bound_Poiss_1}
\varphi(\infty)=0.
\end{equation}
Here $U$ is the applied voltage, and $\varphi_{bi}$ is the
built-in potential.

In addition, the matching conditions for the potential at both
heterointerfaces have to be fulfilled
\begin{equation}\label{cond_Poiss}
\epsilon_{\text{barr}}\frac{d\varphi_{\text{barr}}}{dz}=
\epsilon_{\text{well}}\frac{d\varphi_{\text{well}}}{dz}.
\end{equation}
The indexes "barr" and "well" correspond to the regions of GaAs
barrier and InGaAs quantum well, respectively.

The free carrier concentration $n(z)$ in (\ref{Poisson}) far from
the QW can be calculated as in the case of a homogeneous structure
through the Fermi integral
\begin{equation}\label{Fermi}
n(z)=N_C\frac{2}{\sqrt{\pi}}F_{1/2}\left(-\frac
{E_C-E_F-e\varphi(z)}{kT}\right),
\end{equation}
where $N_{C}$ is the effective density of states in the conduction
band, $E_{F}$ is the Fermi level, $T$ is the temperature, and $k$
is the Boltzmann constant. In contrast, in the vicinity of a
quantum well the carrier concentration should be calculated by
solving Schr\"{o}dinger's equation. The needed spatial
distribution of the electrostatic potential was derived using a
procedure of self-consistency of the Schr\"{o}dinger and Poisson
equations. The essence of the procedure is the sequential
(step-by-step) solution of Schr\"{o}dinger and Poisson equations
until convergence. \cite{Ste72,Ste84,Tan90} As a criterion of
convergence we took an increment of the potential of less than
$10^{-8}$ V in the next iteration.

\begin{figure}
\includegraphics[width=8.5 cm]{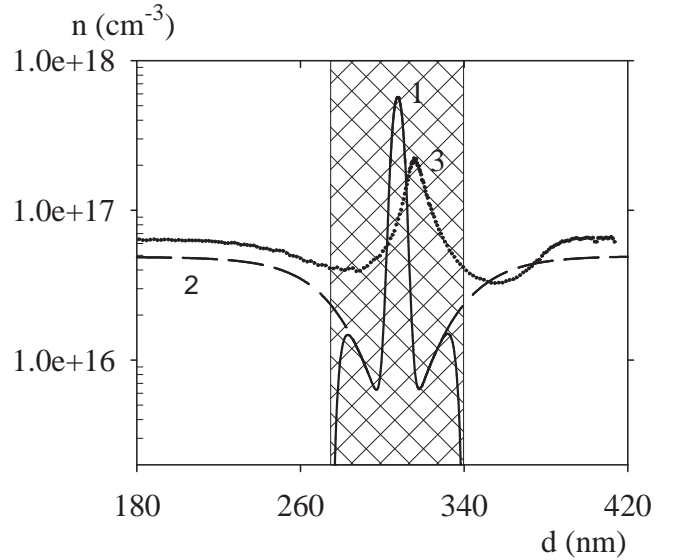}\\
\caption{\label{Fig:true_box} Scheme of computations. The "quantum
box" is shaded. 1 --- Schr\"{o}dinger's concentration; 2 ---
concentration derived from (\ref{Fermi}); 3 --- resulting
calculated "apparent" concentration (matched the experiment).}
\end{figure}

Quantum size effects are important only inside the quantum well
and in its immediate vicinity. So for numerical solution of the
Schr\"{o}dinger equation we used a "quantum box", a narrow region
containing the QW (Fig.~\ref{Fig:true_box}). The optimal width of
the quantum box was chosen experimentally during simulations, in
order to achieve a high precision in determination of the
quantized carrier concentration, and, in balance, to reduce the
computing time. The quantum box width was chosen as nine times the
width $W$ of the corresponding quantum well, which was placed in
the center of the quantum box.

According to the boundary conditions, there must be nodes of the
wave functions at the edges of the quantum box. For this reason
the data calculated close to the quantum box boundaries are
dropped. The length of this region is no more than $1$ $W$, as can
be seen from Fig.~\ref{Fig:true_box}. On the other hand, at
distances about $2-3$ $W$ from the QW the quantization effect is
very weak, and there we can use the Fermi integral (\ref{Fermi})
for deriving the free carrier concentration. The coincidence of
the concentration profiles at this part derived from
quantum-mechanical approach and from (\ref{Fermi}) was used as the
proof for a true solution.

\subsection{\label{Sec:Poisson} Solving the Poisson equation}
The Poisson equation has been solved numerically by Newton's
method relative to the correction term.

The great difference (tenfold) in the values of electron and hole
effective masses in GaAs and nearby ternary InGaAs alloys makes
the Fermi level shift toward the bottom of the conduction band.
Therefore, a significant part of donors (up to 30\%) remains
non-ionized at room temperature, even despite the very low
ionization energy of Si donors in GaAs (about 5 meV \cite{Ada92}).
Because of this, the incomplete donor ionization has to be taken
into account.

To reduce the computation time, it is desirable to use some
approximation of the Fermi integral (\ref{Fermi}). The simplest
exponential approximation is not applicable here because of the
close position of the Fermi level to the bottom of the conduction
band. There is another well known approximation for (\ref{Fermi})
by the expression
\begin{equation}\label{C_n appr}
n(z)=\frac{N_C}{C_{n}+\exp\left(\frac{E_{C}-E_{F}-e\varphi(z)}{kT}\right)},
\end{equation}
that better matches the Fermi integral. The constant $C_n$ here
usually falls between 0.17 and 0.35. \cite{Abo93}

In order to minimize the approximation error and to fulfill the
electroneutrality condition on the right-hand side of the
simulated region (i.e.\ in the GaAs substrate) we used the
following procedure: by solving the electroneutrality equation and
using expression (\ref{Fermi}) the Fermi level position is
determined at $\varphi = 0$. Then equating (\ref{Fermi}) to
(\ref{C_n appr}) one can derive the current adaptivity constant
$C_n$. At another $\varphi$ the maximum relative error of such
approximation does not exceed $3\times10^{-4}$.

The electrostatic potential was written as an initial
approximation ${\varphi}_0(z)$ and a correction term
${\Delta}\varphi(z)$:
\begin{equation}\label{phi_correct}
\varphi(z)=\varphi_0(z)+\Delta\varphi(z).
\end{equation}

To linearize the Poisson equation (\ref{Poisson}) the expression
for $n$ was decomposed into a Taylor series including linear term
relative to the correction ${\Delta}\varphi(z)$.

Then a finite-difference analog of the Poisson equation has been
rewritten as a system of linear equations with a characteristic
three-diagonal form. To get a high precision solution in a
reasonable time different mesh steps were used inside and outside
the quantum box. The number of points in the mesh was 8000,
including about 1500 in the quantum box. The Gauss method was
applied to solve the system with some modifications based on
obvious symmetry of the equations.

After getting the correction ${\Delta}\varphi(z)$ a new potential
was obtained according to (\ref{phi_correct}).

\subsection{\label{Sec:Schrodinger} Solving the Schr\"{o}dinger
equation}
The effective mass, one-dimensional Schr\"{o}dinger
equation can be written as \cite{Ste84}
\begin{equation}\label{Schrod}
-\frac{\hbar^2}{2}\frac{d}{dz}\frac{1}{m^*(z)}\frac{d\psi_i(z)}{dz}+
V(z)\psi_i(z)=E_i\psi_i(z),
\end{equation}
where $E_{i}$ are the eigenvalues, $\psi_i$ are the corresponding
eigenvectors, $m^*$ is the coordinate-dependent electron effective
mass. $V(z)$ is the effective potential energy:
\begin{eqnarray}\label{Vx_Schroed}
V(z)=\left \{
\begin{array}{l l}
e\varphi(z)+{\Delta}E_{C} & \text{-- inside QW,}\\
e\varphi(z) & \text{-- outside QW.}
\end{array}
\right.  
\end{eqnarray}
${\Delta}E_C$ is the conduction band offset.

We used boundary conditions of Neuman's type at the ends of the
quantum box.

The finite-difference analog for (\ref{Schrod}) was obtained using
the three-point formula
\begin{equation}\label{3-pointSchrod}
-\frac{\hbar^2}{2m^{*}_{j}}\frac{\psi_{i,j-1}+\psi_{i,j+1}-2\psi_{i,j}}{h^{2}_{j}}+
V_{j}\psi_{i,j}=E_{i}\psi_{i,j},
\end{equation}
where $j$ identifies the point on the one-dimensional mesh, and
$h_j$ is the distance between the mesh nodes (a step of the mesh).

In addition to boundary conditions, at the heterojunctions the
following matching conditions should be maintained between the
derivative of wave functions inside and outside the quantum well:
\begin{equation}\label{Cond_Schrod_num}
\frac{1}{m^{*}_{\text{barr}}}\frac{\Delta\psi_{\text{barr}}}{h_{\text{barr}}}=
\frac{1}{m^{*}_{\text{well}}}\frac{\Delta\psi_{\text{well}}}{h_{\text{well}}}.
\end{equation}

The Schr\"{o}dinger equation (\ref{Schrod}) was solved numerically
by the well known "shooting" method with some improvements aimed
to reduce the computation time.

The number of points in the mesh should be enough to eliminate the
error due to substitution of the derivative with the
finite-difference approximation (\ref{3-pointSchrod}). We compared
the results of numerical solution of (\ref{Schrod}) by the
shooting method with the well known analytical solution for a
rectangular quantum well. \cite{Lan} It was found that the mesh
size of about 1500 points yields quite good accuracy with a
relative error in eigenvalue determination less than $10^{-3}$ for
almost all levels.

After the set of eigenvalues $E_i$ and corresponding eigenvectors
$\psi_i(z)$ had been obtained, the carrier concentration in the
region of QW was calculated via local density of states from the
expression \cite{Tan90, Bro96}
\begin{equation}\label{n_Well}
n(z)=\frac {m^*(z)kT}{\pi \hbar^2}\sum_i
\ln\left[1+\exp\left(\frac
{E_F-E_i}{kT}\right)\right]|\psi_i(z)|^2,
\end{equation}
using the condition of normalizing the wave functions
\begin{equation}\label{wfun_normaliz}
\int_{-\infty}^{+\infty}|\psi_{i}(z)|^{2}dz=1.
\end{equation}
Summation in (\ref{n_Well}) runs over all subbands.

The prefactor in the expression (\ref{n_Well}) before the square
of wave function is considered as the number of electrons per unit
area in the $i$th - subband.

In the quantum well region the concentrations of both bound and
free electrons were calculated from the Schr\"{o}dinger equation.
We consider this a more correct approach than the simple summation
of bound (inside QW) and free carriers (above QW) used, for
example, in Ref.~\onlinecite{Bro96}. But, due to the finite width
of the quantum box the continuum of free electron states in this
scheme of computations is represented by a set of discrete levels
with energies determined by the size of the quantum box.

\begin{figure}
\includegraphics[width=8.0 cm]{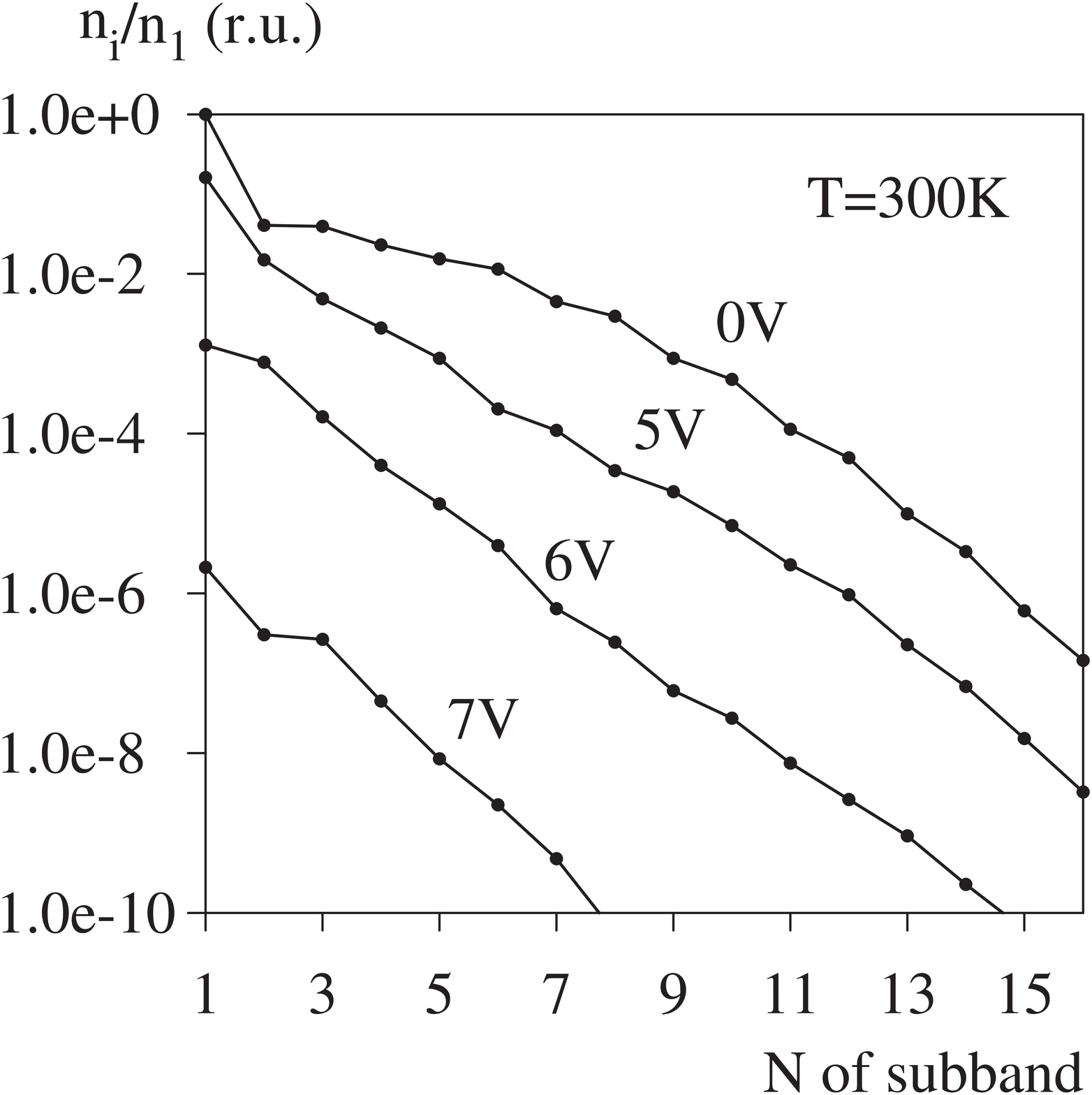}\\
\caption{\label{Fig:Conc_16} Electron concentration in first 16
energy subbbands (relative units) at different reverse biases.
Sample \#300 (${\Delta}E_{C}=175$ meV), T=300K.}
\end{figure}

In order to sum up all charge carriers in the quantum well region
we took into account the 16 lowest energy levels.
Fig.~\ref{Fig:Conc_16} shows the occupation of the 16 energy
levels at different bias. The carrier concentration in the first
subband $n_1$ was at least ten times greater than in the second
one, and the concentration in 16th subband (and all higher
subbands) is below $10^{-7}$ of $n_1$ and can be neglected.

\begin{figure}
\includegraphics[width=8cm]{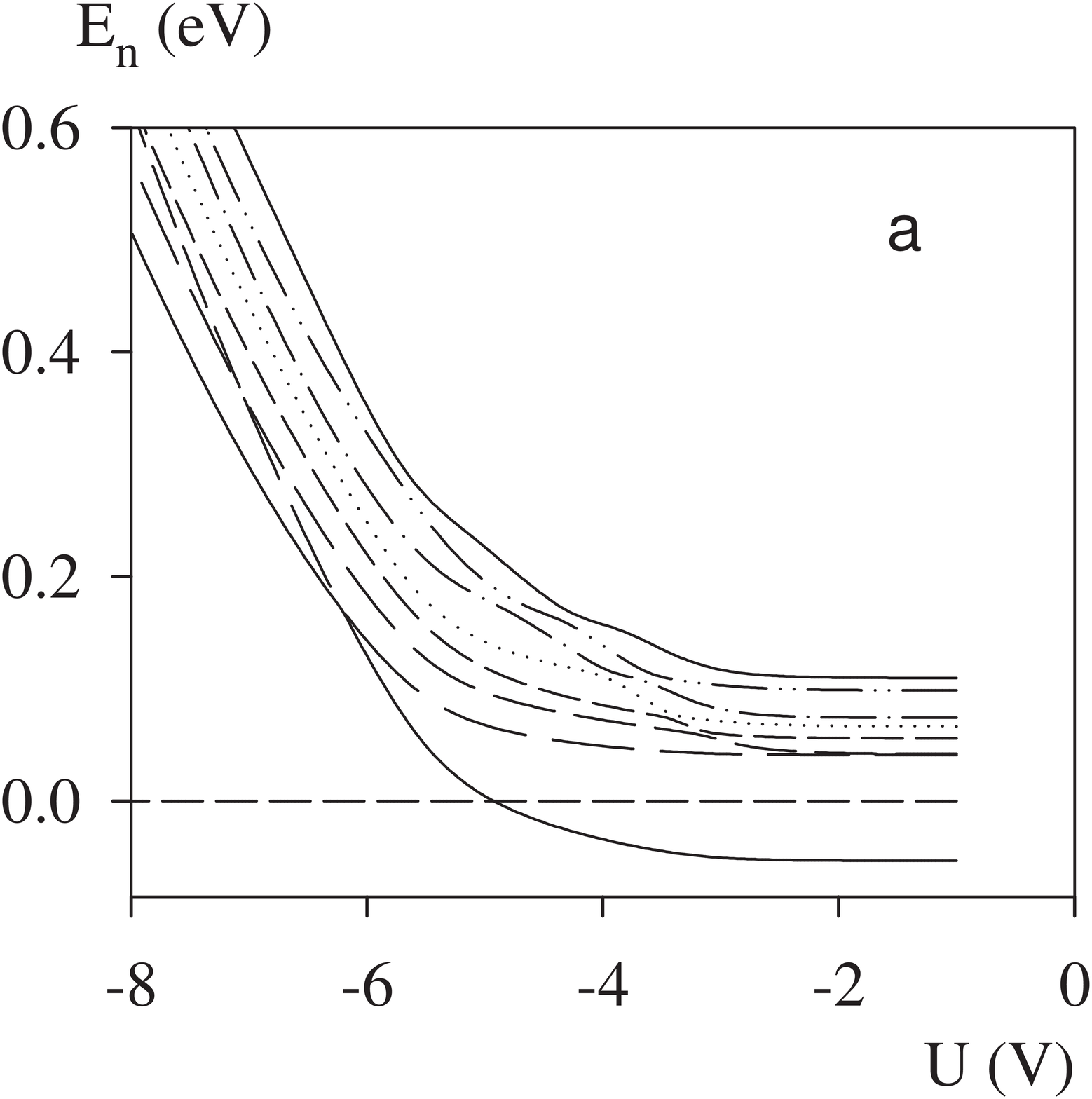}\\
\includegraphics[width=8cm]{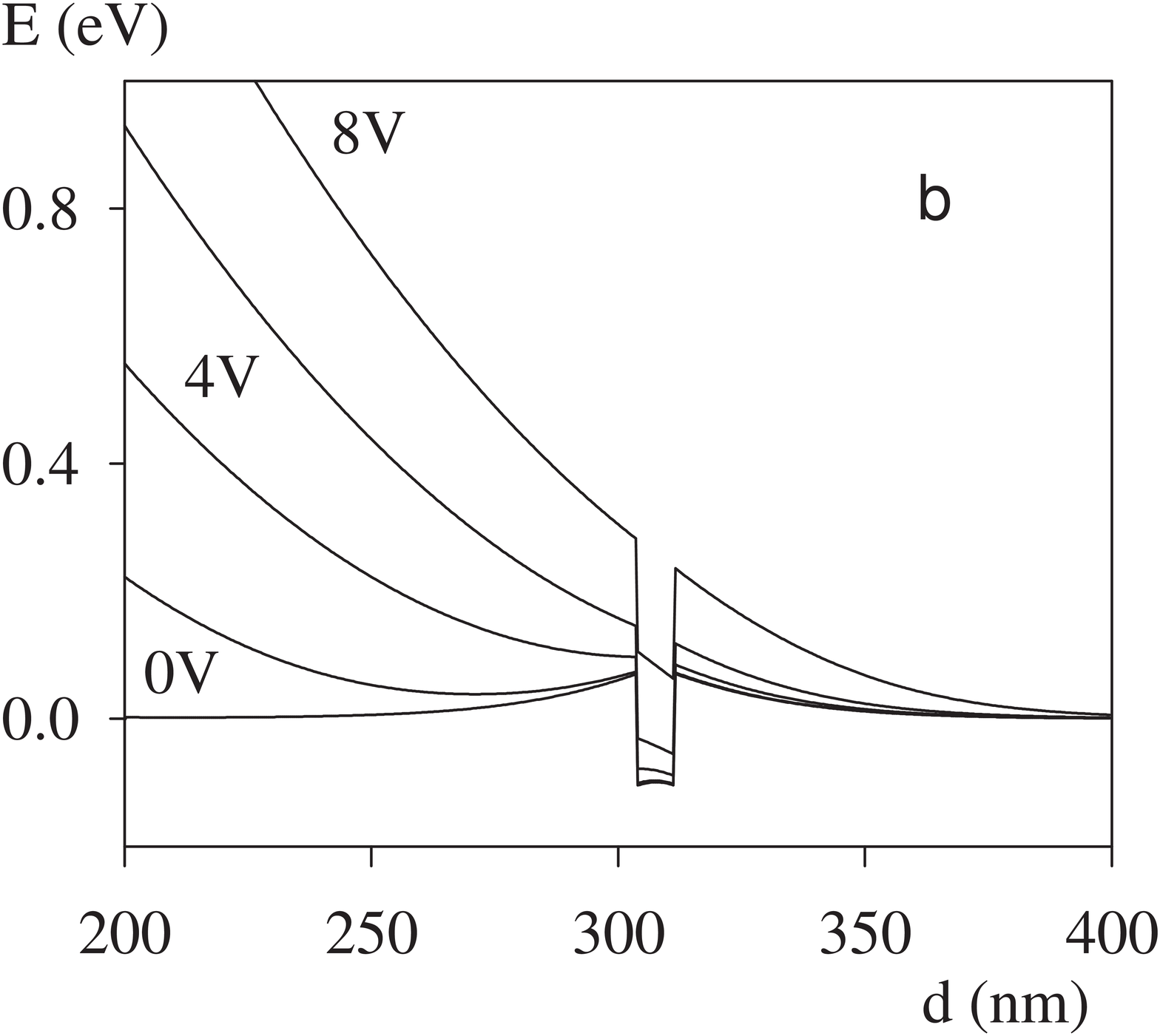}\\
\caption{\label{Fig:Lev300&c-bottom} Results of numerical
calculations for sample \#300 ($x=0.23$; well width = 7.2 nm;
offset ${\Delta}E_{C} = 175$ meV):\\
a) first 8 energy levels as a function of $U_{\text{rev}}$;  b)
the bottom of conduction band near the QW at different
$U_{\text{rev}}$.}
\end{figure}

\begin{figure}
\includegraphics[width=8.5cm]{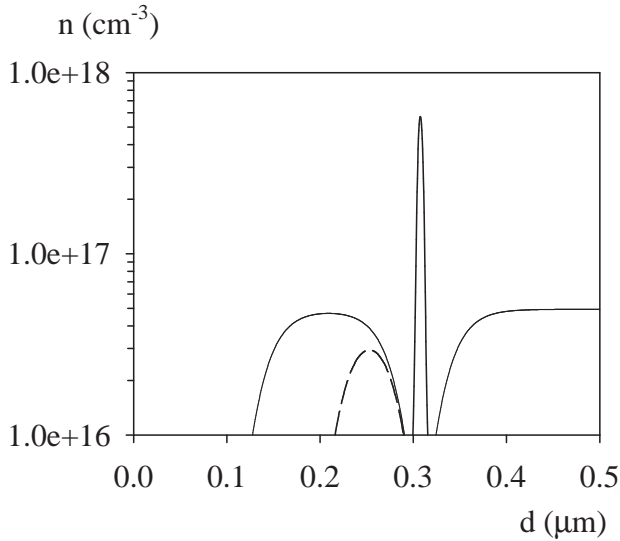}\\
\caption{\label{Fig:simulated profile} Simulated concentration
profiles of electrons in the region of QW at $U=-1$ V (solid) and
$-2.5$ V (dashed line). Sample \#300.}
\end{figure}

The results of the computation of energy states in dependence of
applied reverse bias for the sample \#300 (${\Delta}E_{C}=175$
meV) are shown in Fig.~\ref{Fig:Lev300&c-bottom}(a), and a lineup
of the conduction band bottom for this structure is depicted in
Fig.~\ref{Fig:Lev300&c-bottom}(b). As can be seen, a single bound
level is observed in the structure with an energy of 49 meV in
equilibrium. (The bottom of the conduction band in the
electroneutrality region was taken as zero). Starting
approximately at $-2.5$ V the space charge regions of the Schottky
barrier and the QW merge (Fig.~\ref{Fig:simulated profile}), and
the penetrating electric field bends the conduction band bottom
near the QW, forcing the bound level to lift up. At $U=-5$ V the
level becomes unbound.

To calculate the C-V characteristics so called "quasistatic
approach" \cite{Joh71} was applied. The capacitance of a structure
is the first derivative of the total charge. The latter can be
derived via the flow of electric field across the surface
according to the Gauss theorem. The spatial distribution of
electrostatic potential $\varphi$ is calculated during solution of
the Poisson equation, so one can derive the value of the electric
field at the surface at different applied biases:
\begin{equation}\label{Sur_field}
E_{\text{surf}}=\frac{\varphi_1-\varphi_0}{h_{\text{bulk}}},
\end{equation}
and, hence, build up the capacitance-voltage characteristic (or
restore the apparent concentration profile) using (\ref{n_w}).

\section{\label{Sec:Results} Results of simulation and discussion}
As has been established earlier, \cite{Kro80,Tsc96} there exists a
certain discrepancy between the true and "apparent" concentration
profiles of free charge carriers near a heterojunction, a quantum
well or a quantum dot. \cite{Bru97,Kap99,Zub01} An apparent
profile, obtained in experiment, is more smeared in comparison to
the true one and has a shift in the peak position (see
Fig.~\ref{Fig:true_box}). The general reason for this discrepancy
is the indirect and non-equilibrium procedure of concentration
profile restoration from C-V-measurements. Generally, this
technique involves differentiation of the C-V-curve (\ref{n_w}) in
the approximation of fully depleted space charge region and does
not take into account the problem of Debye smearing. In the case
of QW profiling, where one expects sub-Debye resolution, this
standard technique leads to an essential distortion of the
apparent profile. So, for the goal of adequate fitting, during
simulations we must accomplish just the same procedure of
restoring the apparent profile as in real experiment and,
particularly, the bias voltage increment in the simulation must be
equal to the voltage step used in the experiment.

\begin{figure}
\includegraphics[width=8.5 cm]{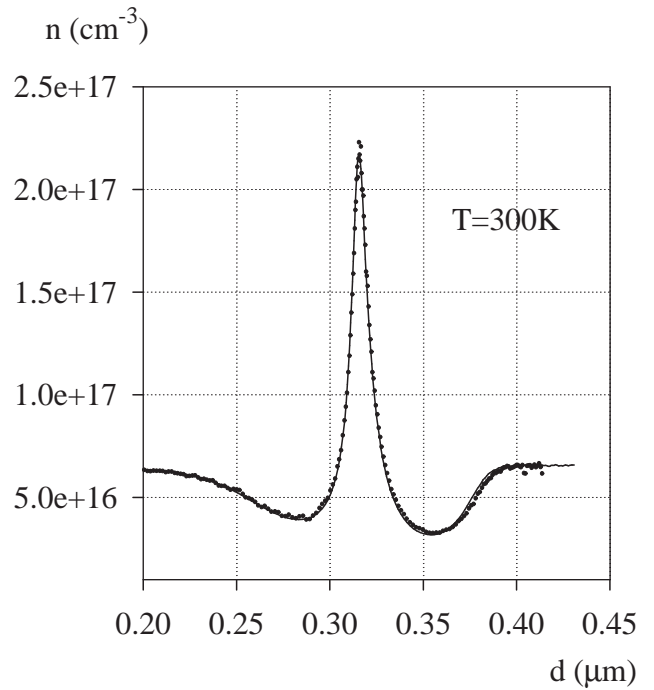}\\
\caption{\label{Fig:fitting300} Experimental (dots) and fitted
(solid) apparent concentration profiles of
In$_{0.23}$Ga$_{0.77}$As/GaAs quantum well. (Sample \#300,
T=300K). Best fit ${\Delta}E_{C} = 175$ meV.}
\end{figure}

\begin{figure}
\includegraphics[width=8.5 cm]{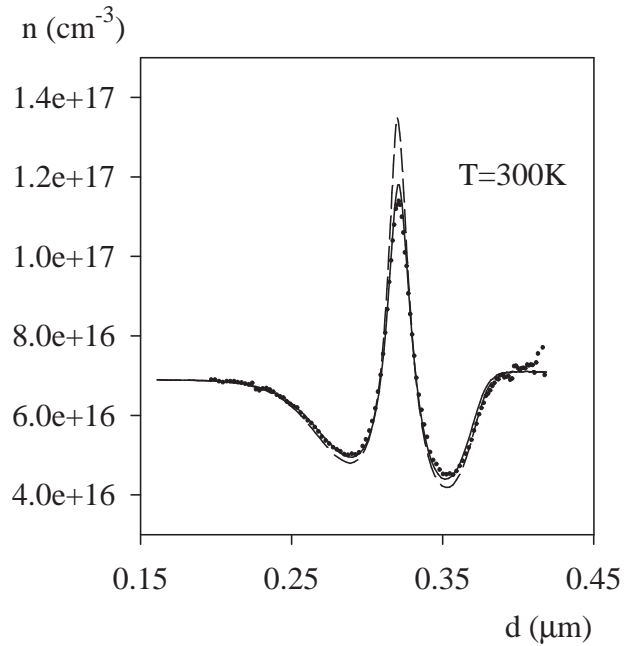}\\
\caption{\label{Fig:fitting303} Resolution of the fitting. The
results for the best fit (${\Delta}E_{C} = 120$ meV, solid line)
and for ${\Delta}E_{C} = 130$ meV (dashed line) are presented.
Dotted curve --- the experimental apparent profile for sample
\#303 ($x = 0.145$).}
\end{figure}

The results of fitting for two samples are presented in
Figs.~\ref{Fig:fitting300} and \ref{Fig:fitting303}. As can be
seen, excellent matching is obtained. This proves the correctness
of the used model. One should underline again that due to the high
quality of the specially fabricated for C-V-measurements samples
no additional adjustable parameters like an impurity concentration
gradient or a charge at the heterojunction had to be used in the
fitting procedure. The only fitting parameter was the conduction
band offset, ${\Delta}E_C$. The value of majority carrier
concentration was taken on the shoulders of the measured
concentration profiles. Parameters for In$_x$Ga$_{1-x}$As, needed
for the calculations, were taken from. \cite{Lev99,Ada92} Fig.
\ref{Fig:fitting303} also demonstrates the resolution of our
fitting. For a medium In-content ($x = 0.14$) the error was less
than 10 meV. It was found that the resolution is approximately
directly proportional to the alloy composition of the quantum
well. In general, we estimate the relative error in the
determination of band offsets as less than 10\% within the
measured range of $x$.

\begin{figure}
\includegraphics[width=8.5 cm] {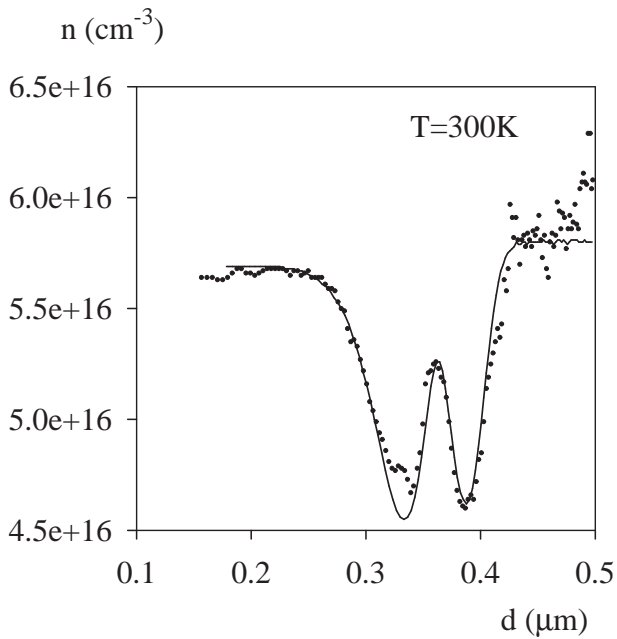}\\
\caption{\label{Fig:Sample298} Apparent concentration profile of
In$_{x}$Ga$_{1-x}$As/GaAs quantum well with low In-content
$x_{\text{In}} = 0.065$ (dotted) and fitted curve (solid) with
$\Delta$E$_C = 55$ meV. (Sample \#298, T=300K).}
\end{figure}

An interesting example of fitting for the sample with the smallest
In-content in QW ($x = 0.065$, sample \#298) is presented in
Fig.~\ref{Fig:Sample298}. Here the apparent peak of enrichment in
the QW is even smaller than the value of the impurity
concentration, despite the spatial confinement inside the QW. The
simulated profile in this case is very sensitive to the band
offset (the error is about 5 meV), however, the fitting is not as
good as for other compositions. For such a weak concentration peak
the presence of residual impurities in the quantum well and in
adjacent spacers begins to play an essential role. One also should
bear in mind for explanation the increased relative value of the
experimental noise.

\begin{figure}
\includegraphics [width=8.0 cm] {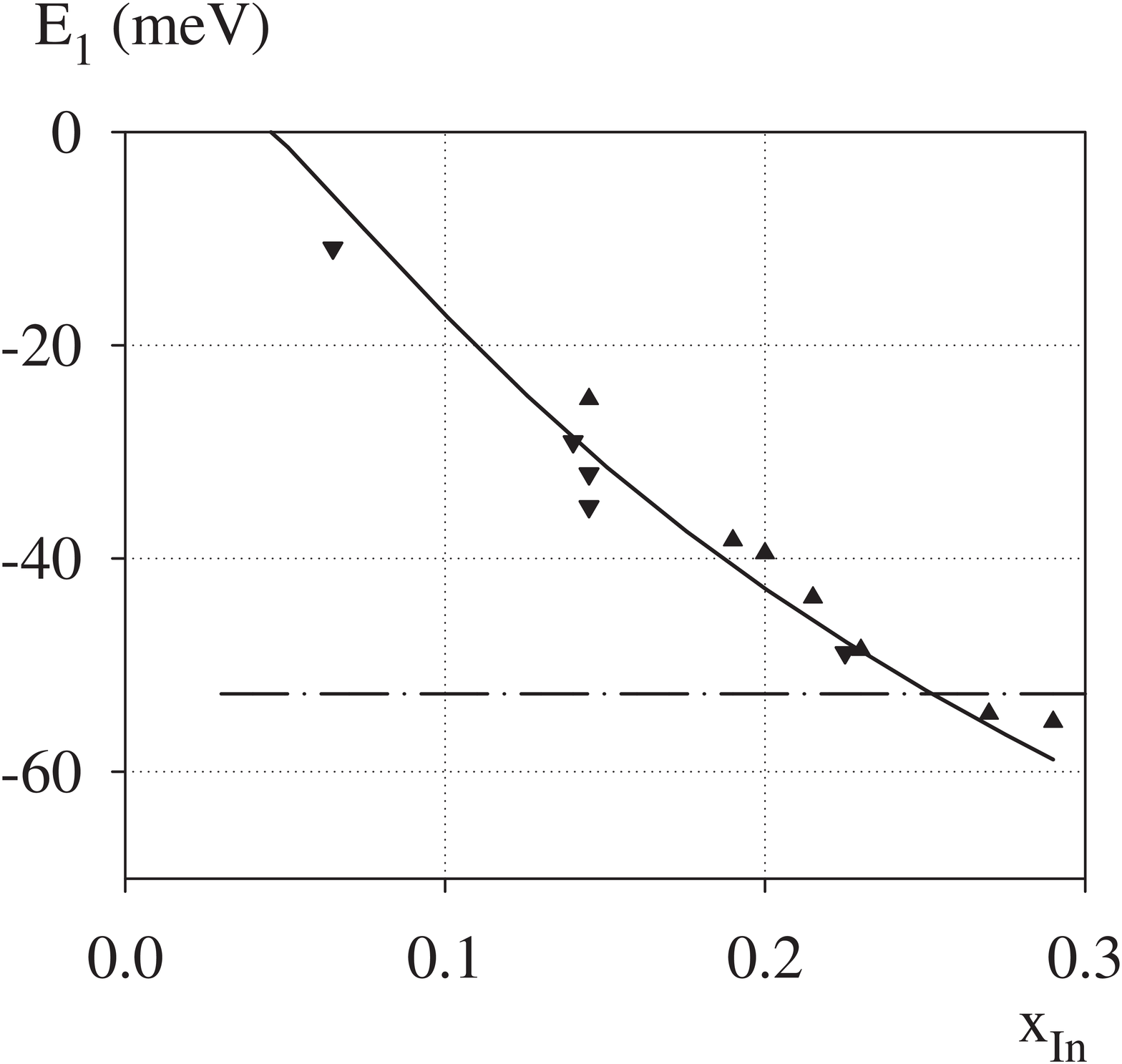}\\
\caption{\label{Fig:Eigens} Position of bound energy level $E_1$
in strained In$_x$Ga$_{1-x}$As/GaAs-QWs (triangles up for
$w\leq7.2$ nm and triangles down for $w>7.2$ nm). Solid line ---
calculated dependence $E_{1}$ of $x_{\text{In}}$ in the assumption
$w = 7.2$ nm and for the doping concentration as in the sample
\#300. Dash-dot line --- the corresponding position of Fermi
level.}
\end{figure}

In Table \ref{tab:Samples} we summarized the conduction band
offsets in strained pseudomorphically grown
In$_x$Ga$_{1-x}$As/GaAs ($0.06 \leqslant x \leqslant 0.29$)
quantum wells obtained in our study. Only one bound level was
observed in all samples. Its depth in equilibrium ($U = 0$ V) is
depicted in Fig.~\ref{Fig:Eigens} as a function of composition.
One can see that for compositions $x<0.25$ the level appears above
the corresponding Fermi level. Nevertheless, the occupation in the
subband remains significant to provide an excess of apparent
carrier concentration in QW region over the dopant value in all
samples, except for $x=0.065$.

From Fig.~\ref{Fig:Eigens} it can be seen that there is no bound
energy level for $x<4$\%. Indeed, the weak doping in the adjacent
to QW spacers leads to an additional conduction band bending near
the QW, which lifts the energy level up. The effect of
disappearing bound level does not exist if the spacers are absent
(at least, down to extremely low $x$, about 1\%, when errors in
numerical calculations begin to occur).

\begin{figure}
\includegraphics [width=8.5cm] {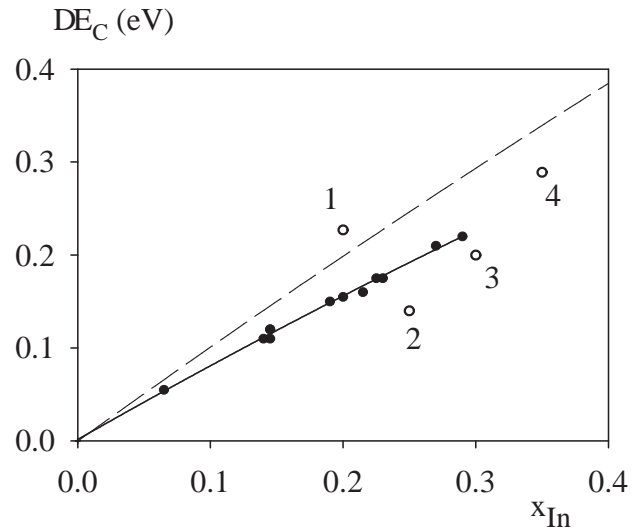}
\caption{\label{Fig:jap2me} Conduction band offsets in strained
In$_x$Ga$_{1-x}$As/GaAs QWs as a function of composition. Dashed
line --- as recommended in Ref.~\onlinecite{Vur01}. Open circles
--- the latest results of capacitance and optical investigations
on strained In$_x$Ga$_{1-x}$As/GaAs--QWs: 1 --
Ref.~\onlinecite{Liw98};  2 -- Ref.~\onlinecite{Let91};  3 --
Ref.~\onlinecite{Kar96};  4 -- Ref.~\onlinecite{Ari00}.}
\end{figure}

In Fig.~\ref{Fig:jap2me} the results on conduction band offsets in
strained In$_x$Ga$_{1-x}$As/GaAs-QWs obtained during the numerical
fitting to the experimental C-V characteristics are presented. The
"recommended" curve from the above mentioned review \cite{Vur01}
is also depicted. The "recommended" values of band offsets are
higher by about 25\% in comparison to our results for strained
quantum wells. The origin of this difference most probably is the
presence of significant degree of elastic strain in pseudomorphic
InGaAs on GaAs. The deformation potential in compressively
strained InGaAs modifies the energy lineups of the
heterostructure. Estimations on the base of model-solid theory
\cite{Wal89} predict an increase of the band gap of compressively
strained InGaAs in comparison to the value $\Delta$E$_g$ for bulk
material (this effect again is of the order of 25\%). So, the
absolute values of band discontinuities are smaller in
compressively strained quantum wells than in relaxed single
heterostructures or thick double heterostructures. Recent results
of other researchers on strained QWs, mainly obtained by
capacitance techniques (also depicted in Fig.~\ref{Fig:jap2me}),
are in reasonable agreement with ours, but exhibit significant
scattering.

The experimentally obtained dependence ${\Delta}E_C = f(x)$ is
close to a straight line with only little bowing. From fitting the
curve to a parabola we propose the expression ${\Delta}E_{C}(x) =
0.814x - 0.21x^{2}$ for the conduction band offsets in strained
In$_x$Ga$_{1-x}$As/GaAs quantum wells in the composition range $0
< x < 0.3$.

\section{Summary}
Aiming to get accurate and precise values for conduction band
offsets, a set of high quality samples containing strained
In$_x$Ga$_{1-x}$As/GaAs quantum wells was grown in the composition
range  $0.06 \leqslant x \leqslant 0.29$. Specially for
C-V-measurements a constant impurity concentration in the cladding
layers was maintained during the growth in order to eliminate
uncertainties in subsequent numerical simulations. A fitting
procedure of experimentally obtained apparent concentration
profiles has been implemented using self-consistent solution of
Schr\"{o}dinger and Poisson equations. All important information
about the properties of the quantum well structures was derived:
the majority carrier profiles, the positions of energy levels,
corresponding wave functions, profile of the conduction band
bottom, as well as the dependencies of the above mentioned
parameters on the applied electric field. The presence of only one
bound level was discovered in all samples. The conduction band
offsets in strained In$_x$Ga$_{1-x}$As/GaAs quantum wells follow
the expression ${\Delta}E_{C}(x) = 0.814x - 0.21x^{2}$.

\bibliography{cv_prb}
\end{document}